\documentclass[10pt,prl,aps,twocolumn,superscriptaddress,showpacs,floatfix]{revtex4}
\usepackage{graphicx}

\usepackage{amssymb}

\begin{document}

\title{Symmetry breaking and criticality in tensor-product states}

\author{Chen Liu}
\affiliation{Department of Physics, Boston University, 590 Commonwealth Avenue, Boston, Massachusetts 02215, USA}

\author{Ling Wang}
\affiliation{Institut der Theoretischen Physik, Universit\"at Wien, Boltzmanngasse 3, A-1090 Vienna, Austria}

\author{Anders W. Sandvik}
\affiliation{Department of Physics, Boston University, 590 Commonwealth Avenue, Boston, Massachusetts 02215, USA}

\author{Yu-Cheng Su}
\affiliation{Department of Physics and Center of Quantum Science and Engineering, National Taiwan University, Taipei, Taiwan}

\author{Ying-Jer Kao}
\affiliation{Department of Physics and Center of Quantum Science and Engineering, National Taiwan University, Taipei, Taiwan}

\begin{abstract}
We discuss variationally optimized matrix-product states for the transverse-field Ising chain, using $D\times D$ matrices with 
small $D \in \{2-10\}$. For finite system size $N$ there are energy minimums for symmetric as well as symmetry-broken states, which 
cross each other at a field value $h_c(N,D)$; thus the transition is first-order. A continuous transition develops as $N\to\infty$. The 
asymptotic critical behavior is then always of mean-field type (the magnetization exponent $\beta=1/2$), but a window of field strengths where 
true Ising scaling holds ($\beta=1/8$) emerges with increasing $D$. We also demonstrate asymptotic mean-field behavior for 
infinite-size two-dimensional tensor-product (iPEPS) states with small tensors.
\end{abstract}

\date{\today}


\maketitle

Methods based on matrix-product states (MPSs) \cite{aklt,ostlund} have become the primary computational tools for studies of static 
as well as dynamic properties of one-dimensional quantum many-body systems \cite{verstraete1}. Key steps in the development of these 
methods were White's density matrix renormalization group (DMRG) \cite{white,schollwock}, the demonstration by \"Ostlund and Romer
of its connection to MPSs \cite{ostlund}, and later important insights from the field of quantum information theory. In particular,
the concept of entanglement entropy (the area law) both explains the success of the approach in one dimension and its failure 
(violation of the area law) in higher dimensions \cite{verstraete2,verstraete3,hastings}. The formulation of computational methods 
directly in terms of MPSs also led to a framework for efficient optimization of these states independently of the DMRG method
\cite{murg,vidal1,mcculloch}, and to a long-sought way of computing time evolution \cite{vidal2}. The MPS approach also has a natural 
extension to higher dimensions which {\it does} obey the area law \cite{verstraete3}---tensor-product states; also referred to as 
projected-entangled-pair-states (PEPSs) \cite{nishino,verstraete4}.

In spite of numerous successful applications of MPS-based methods, some fundamental aspects of this class of quantum states have 
not yet been studied in detail. It is well known that the finite size $D$ of the $D \times D$ matrices (the elements of which are
the variational parameters) imposes a finite correlation length, and recently it has been recognized that scaling in $D$ for infinite 
system size $N$ can be carried out as an alternative to finite-size scaling \cite{tagliacozzo} (i.e., $D$ and $N$ can be considered 
as different but equally valid ways to regularize the calculations). As in mean-field theory (which corresponds to $D=1$), an MPS can 
break symmetries of the hamiltonian at a phase transition.
Exactly how the critical behavior of the order parameter (the true scaling exponent $\beta$) emerges as a function of $N$ and $D$ 
has not been studied systematically, however. This may be partially due to technical challenges in properly optimizing an MPS 
close to a phase transition. Such issues are present also for the PEPS approach in two dimensions. Order-parameter curves often 
exhibit rounding \cite{orus}, that may appear due to incomplete convergence, approximations made \cite{tagliacozzo}, 
or due to external fields included to stabilize the calculation \cite{nagaj}. Nevertheless, the behavior slightly away from the 
transition can be well described by the expected critical exponent \cite{tagliacozzo,jordan,gu}. The question remains 
{\it whether this is the true critical behavior of the MPS or PEPS  variational ansatz with finite D}, or whether there could
eventually be a cross-over to a different asymptotic form. 

In this {\it Letter} we study the asymptotic critical behavior by using numerically stable high-precision optimization 
methods for small $D$, for both finite and infinite $N$. Using the transverse-field Ising model as a demonstration, we show that 
access to the true critical behavior of an MPS requires very high numerical precision; in some cases higher than the double-precision 
(64-bit) floating point arithmetic normally used. Optimizing the states to the required precision, we show that the asymptotic 
critical behavior of the order parameter is always mean-field like ($\beta=1/2$). The true universal exponent for the model 
($\beta=1/8$) emerges in a window which approaches the critical point as $D$ increases. We also shows results in two dimensions 
for an infinite-size PEPS (iPEPS), optimized using a recently developed numerically stable scheme \cite{wang1}. Also here we find 
$\beta=1/2$ asymptotically.

\begin{figure}
\centerline{\includegraphics[width=8cm, clip]{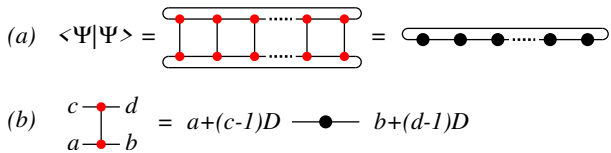}}
\vskip-1mm
\caption{(Color online) (a) The norm of an MPS expressed as the contraction of a tensor network. Carrying out the summations
over the spin indices first (vertical bonds), as indicated in (b), gives a simple trace of a product of matrices of size 
$D^2 \times D^2$ (with a possible labeling of the elements indicated).}
\label{fig1}
\vskip-3mm
\end{figure}

First, consider the simplest kind of MPS for a periodic, translationally invariant $S=1/2$ spin chain;
\begin{equation}
|\Psi\rangle = \sum_{\{ \sigma^z\}} {\rm Tr}\{ A(\sigma^z_1)A(\sigma^z_2)\cdots A(\sigma^z_N)\}|\sigma^z_1,\ldots,\sigma^z_N\rangle,
\end{equation}
where $\sigma^z_i=\pm 1$ and $A(\pm 1)$ are two hermitian $D\times D$ matrices. As illustrated in Fig.~\ref{fig1}, the normalization
of this state can be expressed as the contraction of a network of $3$-index tensors $A_{ab}(\sigma)$, where $\sigma=\pm 1$
is the physical index. By contracting over the physical indices first, matrices $B$ of size $D^2\times D^2$ are obtained;
\begin{equation}
B_{ij} = A_{ab}(+1)A^*_{cd}(+1) + A_{ab}(-1)A^*_{cd}(-1),
\label{bdef}
\end{equation}
where $i=a+(c-1)D$ and $j=b+(d-1)D$. The normalization is then simply
\begin{equation}
\langle \Psi|\Psi\rangle = {\rm Tr}\{B^N\}.
\label{norm}
\end{equation}
Expectation values can be computed in a very similar way, with some of the $B$ matrices in the product replaced by the matrix 
obtained as in (\ref{bdef}) but with the operator in question first acting on the physical index \cite{verstraete1}. 
For instance, the magnetization $m$ is given by
\begin{equation}
m=\langle \sigma^z_i \rangle = \frac{{\rm Tr}\{MB^{N-1}\}}{{\rm Tr}\{B^{N}\}},
\end{equation}
where the matrix $M$ is
\begin{equation}
M_{ij} = A_{ab}(+1)A^*_{cd}(+1) - A_{ab}(-1)A^*_{cd}(-1).
\label{mdef}
\end{equation}
The generalization to expectation values of products of two or more operators is straight-forward.

The matrix $B$ is exactly analogous to the transfer matrix in classical statistical mechanics. With $U$ the unitary matrix that 
diagonalizes $B$, giving its eigenvalues $\lambda_1,\ldots,\lambda_{D^2}$, the magnetization can be written as
\begin{equation}
m= \frac{\sum_i [U^{-1}MU]_{ii}\lambda^{N-1}_i}{\sum_i\lambda^N_i}.
\label{mzexpn}
\end{equation}
As in the transfer-matrix approach, the $N\to \infty$ limit can be taken by keeping only the leading eigenvalue; assumed here 
to be $\lambda_1$. The magnetization is then
\begin{equation}
m= \frac{1}{\lambda_1}\sum_{i,j} v_{1i}^*v_{1j}M_{ij},
\label{mzexp}
\end{equation}
where $v_1$ is the eigenvector of $B$ corresponding to $\lambda_1$.

Given a hamiltonian $H$, the problem is how to find the matrices $A(\pm 1)$, of given size $D$, that best reproduce the ground state. 
This can be formulated as a variational problem; to minimize the energy $E=\langle \Psi|H|\Psi\rangle$. Several different 
optimization methods have been developed. For finite $N$, the translational invariance is typically broken as a series of 
local optimizations are carried out, sweeping back and forth through an open chain \cite{verstraete1} (similar to DMRG calculations 
\cite{white,schollwock}). In a periodic chain, where the calculation is more 
demanding, uniformity is gradually restored as the matrices converge. For $N=\infty$, the most efficient approach is Vidal's time 
evolving block decimation (TEBD) scheme \cite{vidal1}, where the ground state is projected out in the limit of long imaginary time 
\cite{verstraete1,mcculloch}, starting from an initial (e.g., random) state. Similar methods have also been developed for 
two-dimensional iPEPSs, where expectation values cannot simply be expressed in eigenvalue forms such as (\ref{mzexpn}), but 
where good approximations to the contractions can still be defined and evaluated using TEBD-like methods \cite{jordan}.

\begin{figure}
\centerline{\includegraphics[width=8cm, clip]{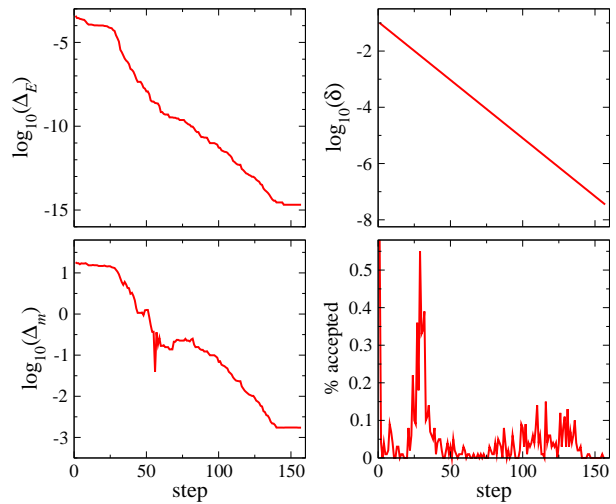}}
\vskip-2mm
\caption{(Color online) Stochastic energy minimization with $10^4$ updates per step for a $D=4$ MPS at $h/J=1.01432$, using
64-bit floating point arithmetic. The relative energy and magnetization errors are defined as $\Delta_E=(E-E_{\infty})/E_\infty$, 
$\Delta_m=||m|-|m_{\infty}||/|m_\infty|$, where the subscript $\infty$ refers to results converged at the 128-bit level.}
\label{fig2}
\vskip-3mm
\end{figure}

Here we investigate symmetry breaking and critical scaling of the order parameter in the transverse-field Ising model. 
In one dimension the hamiltonian is
\begin{equation}
H = - J\sum_{i=1}^N \sigma_{i}^z\sigma_{i+1}^z  - h\sum_{i=1}^N \sigma_{i}^x,
\end{equation}
with periodic boundary conditions. This model is exactly solvable \cite{schultz} and has a paramagnetic--magnetic ($m \not =0$) 
transition at $h_c/J=1$. In two dimension, the critical point has been determined using quantum Monte Carlo calculations, giving 
$h_c/J \approx 3.044$ \cite{rieger}. Single-spin mean-field theory ($D=1$, $N=\infty$) gives $h_c/J=2$ and $4$ in one and two dimensions, 
respectively, and the mean-field form of the magnetization is $m\sim (h_c-h)^\beta$ for $h<h_c$, with $\beta=1/2$. The exact critical exponent 
is $\beta=1/8$ in one dimension and $\beta \approx 0.325$ in two dimensions.

\begin{figure}
\centerline{\includegraphics[width=7.25cm, clip]{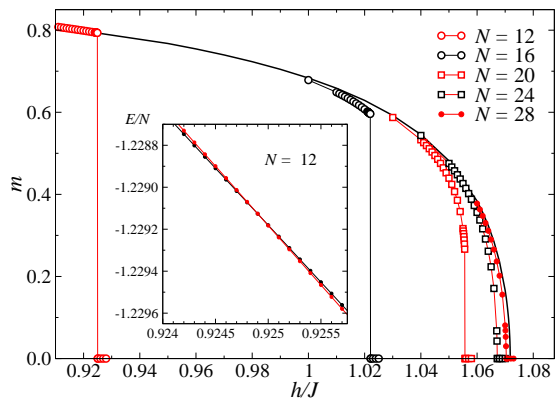}}
\vskip-2mm
\caption{(Color online) Magnetization curves for $D=2$ and different system sizes. The solid curve is for $N=\infty$. The 
inset shows two almost degenerate energy minimums for $N=12$, which cross each other at the transition.}
\label{fig3}
\vskip-3mm
\end{figure}

Considering first MPSs, we optimize the $A$ matrices (which we take as real and symmetric) using two different 
stochastic schemes; one using derivatives and one using only the energy. While the convergence is very slow for large $D$ compared 
to state-of-the-art TEBD \cite{mcculloch}, the methods do not rely on any approximations and are numerically stable. With stochastic
updates, we can avoid potential local minimums in a complex energy landscape. The derivative-based method is the one used in 
\cite{sandvik}, but with exactly computed energies and derivatives for finite $N$. For $N=\infty$, we instead use a brute-force 
scheme with completely random simultaneous updates of all the matrix elements (but keeping the matrices symmetric); 
$A_{ab}(\sigma) \to A_{ab}(\sigma) +\delta[1/2-r_{ab}(\sigma)]$, with uniformly distributed random numbers $r_{ab}(\sigma) \in [0,1)$. 
An update is accepted only if the energy decreases, and then the matrices are normalized so that the largest $|A_{ab}(\sigma)|=1$. 
One step of this procedure typically involves $n \sim 10^3-10^4$ trials. If the acceptance rate is below $10\%$ 
we reduce $\delta$ by dividing by, e.g., $1.1$. To ensure full convergence, when $\delta$ has reached the limit where the updates no 
longer can influence the energy (within the numerical precision), it is reset to a larger value and the process is repeated 
(several times, until no updates are accepted).

Fig.~\ref{fig2} illustrates the brute-force procedure for a $D=4$ MPS optimized at $h/J=1.01432$. The evolution of the errors 
of the energy and the magnetization is shown, along with $\delta$ and the acceptance rate. In this case the acceptance rate 
was always below $10\%$, and $\delta$ therefore decreases after each step. This calculation was carried out using standard 
64-bit floating-point arithmetic, which is reflected in the convergence of the energy to within a relative error of $\approx 10^{-15}$. 
The computation was continued with 128-bit arithmetic, until the energy was converged 
to $\approx 10^{-25}$. The errors graphed in the figure are with respect to this second optimization. The 64-bit optimization took 
only a few minutes, whereas the subsequent 128-bit run took many hours. The computational effort increases very rapidly with $D$, 
and we have only carried out systematic studies up to $D=10$ (for which some points required several weeks of CPU time) \cite{note}.

\begin{figure}
\centerline{\includegraphics[width=7.5cm, clip]{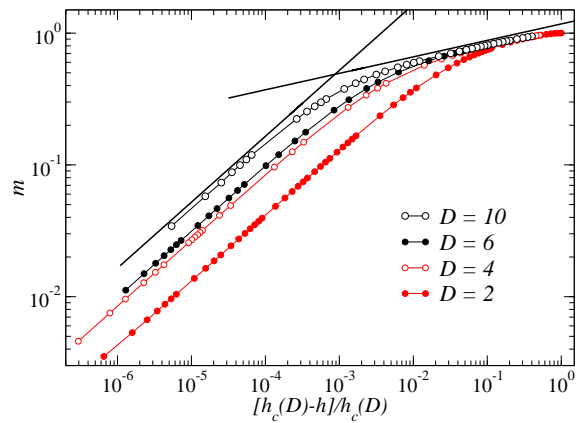}}
\vskip-2mm
\caption{(Color online) Demonstration of asymptotic MPS mean-field behavior and scaling cross-over in one dimension.
The $D$-dependent critical fields are: $h_c/J=1.0717967$ ($D=2$), $1.0143343$ ($D=4$), $1.0063523$ ($D=6$) $1.0021654$ ($D=10$). 
The lines have slopes $\beta=1/8$ and $1/2$.}
\label{fig4}
\vskip-3mm
\end{figure}

Note that while the energy in Fig.~\ref{fig2} has converged to full 64-bit precision, the relative magnetization error is much larger, 
$\Delta_m \approx 10^{-3}$. Using 128-bit arithmetic gives $m=0.031814167$ (where all digits shown are converged). It is well known that
the energy in MPS and DMRG calculations converges much faster than other quantities \cite{schollwock}, but $N=\infty$ results close to 
the critical point appear to be even more sensitive to extremely small energy variations than had been previously anticipated. 
When trying to extract the asymptotic critical behavior of $m$, the problem is accentuated by the fact that it is the relative, not 
absolute error that is relevant. All results to be discussed below have been converged to the level required for a reliable scaling 
analysis.

As shown in Fig.~\ref{fig3}, for finite $N$ the phase transition occurs with sharp magnetization jumps for small $N$, which
become less pronounced as $N$ increases and the transition moves toward higher $h$. The curves converge toward the continuous transition
obtained in the infinite-$N$ calculation. The first-order behavior can be traced to the presence of two energy minimums (shown
in Fig.~\ref{fig3} for $N=12$), which we can track using steepest-decent optimizations starting from large and small 
$h$ (changing $h$ slowly). The diminishing discontinuity with increasing $N$ implies that the minimums move closer to each 
other in parameter space, coinciding at $h_c$ for $N=\infty$. For fixed finite $N$, the discontinuous jumps move toward 
$h=0$ with increasing $D$, reflecting the fact that when $D\to \infty$ an MPS can reproduce the exact spin-inversion symmetric 
($m=0$) ground state of a finite chain.

For $N=\infty$ and any $D$, the optimal state is symmetry-broken below some $h_c(D)$, with $h_c(D)/J \to 1$ as $D \to \infty$. 
The $D$ dependence is not smooth, as has been pointed out before \cite{tagliacozzo}. Here we focus on the behavior of $m$ for 
$h \to h_c(D)$. Thanks to our high-precision data, we can extract $h_c(D)$ reliably using a power-law assumption; 
$m \propto (h_c-h)^\beta$ for $0<m\ll 1$. This always gives $\beta\approx 0.50$ for the best fit, suggesting that the MPS procedure
leads to mean-field behavior for any finite $D$. As shown in Fig.~\ref{fig4}, the true critical behavior ($\beta=1/8$) 
emerges within a window of $h$-values with increasing $D$, with the cross-over to $\beta=1/2$ gradually moving toward $h_c$.

It is perhaps not surprising, after all, that a finite-$D$ MPS cannot reproduce a non-trivial critical exponent asymptotically, 
because the correlation length is finite. Criticality (which can be non-mean-field) in a one-dimensional classical Ising model 
requires long-range interactions \cite{anderson} and the partition function then does not correspond to an MPS with finite $D$. It 
has also been proved that a finite-$D$ MPS can be renormalized to a product state \cite{verstraete5}. It is, however, 
remarkable that the system is so sensitive to incomplete optimization that the asymptotic mean-field behavior of the order 
parameter had not been noted in previous studies \cite{tagliacozzo,nagaj}.

\begin{figure}
\centerline{\includegraphics[width=7.5cm, clip]{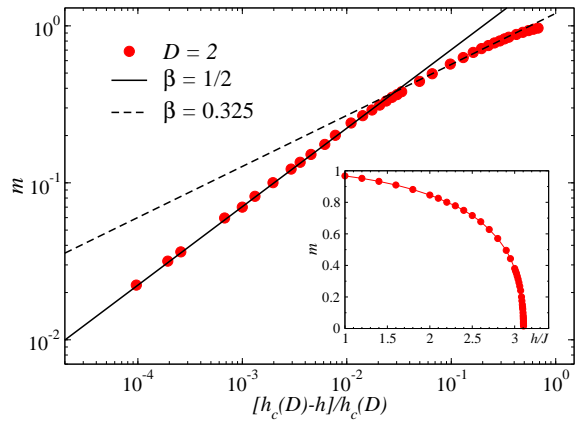}}
\vskip-2mm
\caption{(Color online) Field dependence of the magnetization computed with a $D=2$ iPEPS in two dimensions. The
critical field is $h_c/J=3.1041$.}
\label{fig5}
\vskip-3mm
\end{figure}

We now turn to the two dimensional iPEPS. Non-trivial criticality has been anticipated in this case, even for finite $D$ 
\cite{verstraete3}, because partition functions of classical models with critical points can be written as tensor 
products \cite{nishino}. Magnetization curves closely following the expected power-law with $\beta \approx 0.325$ have
been reported \cite{jordan,orus,gu}, but the calculations are not very accurate close to the critical point. 
We have used an improved iPEPS contraction scheme \cite{wang1}, which is less affected by approximations. 
Fig.~\ref{fig5} shows transverse-field Ising results for $D=2$. An asymptotic mean-field behavior is seen 
unambiguously, and  further away from the critical point there is again a cross-over to a behavior matching closer the true 
$\beta$. However, for $D=2$ the cross-over takes place where $m$ is already large, $\approx 0.5$, and this is not 
actual critical behavior. Unless $D$ is much larger, there will be no clear-cut scaling with the correct exponent.

Our study shows that great care has to be taken when extracting critical scaling forms from the order parameter in MPS
and PEPS calculations. Asymptotic mean-field behavior should be expected, not just for the Ising models considered here, but at
symmetry-breaking transitions in general. This information helps to accurately locate the critical point for small $D$. To extract 
the true critical behavior, it is necessary to carefully examine the behavior for increasing $D$.

{\it Acknowledgments---}We would like to thank P.-C. Chen, I. McCulloch, D. Perez-Garcia, and F. Verstraete for useful discussions, 
and I. McCulloch also for providing TEBD results for comparisons \cite{note}. AWS is supported by NSF grant No.~DMR-0803510 and 
would also like to thank the NCTS of Taiwan for hospitality and funding during a visit. YJK is supported by NCTS and NSC of 
Taiwan under grants NSC 97-2628-M-002-011-MY3 and NTU 98R0066-65, -68.

\null

\end{document}